\documentclass[prl
,twocolumn%
,superscriptaddress%
,showpacs]{revtex4}
\usepackage{graphicx}%
\usepackage{dcolumn}
\usepackage{amsmath}

\makeatletter
\def\btt#1{\texttt{\@backslashchar#1}}%
\DeclareRobustCommand\bblash{\btt{\@backslashchar}}%
\makeatother

\begin{document}
\bibliographystyle{apsrev}
\title{\textit{Ab initio} studies of the tunneling magneto-Seebeck effect: influence of magnetic material} 
\author{Christian Heiliger}%
 \email{christian.heiliger@physik.uni-giessen.de}
\affiliation{%
I. Physikalisches Institut, Justus Liebig University, Giessen, Germany
}
\author{Christian Franz}%
\affiliation{%
I. Physikalisches Institut, Justus Liebig University, Giessen, Germany
}
\author{Michael Czerner}%
\affiliation{%
I. Physikalisches Institut, Justus Liebig University, Giessen, Germany
}
%
\date{\today}
\begin{abstract}
We found a strong influence of the composition of the magnetic material on the temperature dependence of the tunneling magneto-Seebeck effect in $MgO$ based tunnel junctions. We use \textit{ab initio} alloy theory to consider different $Fe_xCo_{1-x}$ alloys for the ferromagnetic layer. Even a small change of the composition leads to strong changes in the magnitude or even in the sign of the tunneling magneto-Seebeck effect. This can explain differences between recent experimental results. In addition, changing the barrier thickness from six to ten monolayers of $MgO$ leads also to a non-trivial change of the temperature dependence. Our results emphasize that the tunneling magneto-Seebeck effect depends very crucially and is very sensitive to material parameters and show that further experimental and theoretical investigations are necessary.
\end{abstract}
%
\maketitle
%
%
%
The recently theoretically predicted \cite{czerner11} and experimentally confirmed \cite{walter11,liebing11} tunneling magneto-Seebeck (TMS) effect in $MgO$ based tunnel junctions belongs to the new field of spin caloritronics \cite{bauer10,bauer12}. In this field the spin-dependent charge transport is combined with energy or heat transport. This means that the spin degree of freedom is exploited in thermoelectrics \cite{silsbee87}. Besides the TMS currently investigated effects in the field of spin caloritronics are the spin-Seebeck effect \cite{uchida08,xiao10}, the
magneto-Seebeck effect in metallic multilayers \cite{gravier06},
the thermal spin-transfer torque \cite{jia11},
the spin-dependent Seebeck effect \cite{lebreton11},
thermally excited spin-currents \cite{tsyplyatyev06},
and the magneto-Peltier cooling \cite{hatami09}. 
In this letter we investigate the role of the ferromagnetic lead material on the TMS. We show that not only the temperature dependence but even the sign of the TMS effect depends crucially on the material composition. 

The TMS effect is the change of the Seebeck coefficient with a change of the magnetic orientation of the ferromagnetic leads relative to each other in a tunnel junction. The size of the effect is given by the TMS ratio
\begin{equation}
\frac{S^P-S^{AP}}{\min (|S^P|,|S^{AP}|)},
\label{eq:TMS}
\end{equation}
where $S^P$ ($S^{AP}$) is the Seebeck coefficient for parallel (anti-parallel) magnetic orientation of the ferromagnetic leads. Therefore, the TMS effect is similar to the tunneling magnetoresistance effect (TMR) \cite{moodera95,miyazaki95}, where one considers the change of the electrical resistance with a change of the magnetic orientation. Note, that in contrast to the resistance the Seebeck coefficient can be positive or negative. Hence, the TMS ratio can have divergences whenever one of the Seebeck coefficients in Eq.~(\ref{eq:TMS}) is zero.

The tunnel junctions we study consist of a $Fe_xCo_{1-x}(001)/MgO/Fe_xCo_{1-x}(001)$ structure embedded between semi-infinite leads. These leads are just acting as reservoirs and are modeled by $Cu$ in the bcc-$Fe$ structure. For the $Fe_xCo_{1-x}$ alloy we use a fixed lattice constant of $0.287 nm$ for all compositions. The thickness of both ferromagnetic leads is 20 monolayers and only symmetric junctions are considered. For the $MgO$ barrier we use six and ten monolayers. Like in our previous studies \cite{czerner11,walter11} we use the ideal positions for the interface layers. In particular, no relaxation effects at the $Fe_xCo_{1-x}/MgO$ interface are taken into account. That way, we really focus on the influence of the change of the electronic structure in the lead material by alloying. Although in most cases, e.g. by sputtering techniques, $Fe_xCo_{1-x}$ is only stable in bcc structure for $x>0.3$ \cite{bonell12} we consider the whole concentration range. Actually, with methods like molecular beam epitaxy it is also possible to grow pure Co leads in bcc-structure \cite{yuasa06}.

For the description of the $Fe_xCo_{1-x}$ alloy we employ the coherent potential approximation (CPA) \cite{butler85,zabloudil05} recently implemented in our Korringa-Kohn-Rostoker (KKR) method~\cite{czerner13}. Within the CPA the alloy is described by an effective medium, which is calculated self-consistently. For the description of the transport properties so-called vertex corrections are essential~\cite{velicky69,carva06,ke08}. The CPA together with the vertex corrections leads basically to the same result as the supercell approach, where one has to average the transport properties over a larger number of different supercells~\cite{czerner13}. In comparison to the supercell approach, the advantage of the CPA is a lower computational effort and the possibility to use an arbitrary composition.
The transport coefficients, in particular the transmission function $T(E)$, are calculated using the non-equilibrium Green's function formalism implemented in the KKR method~\cite{heiliger08} including vertex corrections for the CPA~\cite{czerner13}. From $T(E)$ we calculate the moments
\begin{equation}
L_n=\frac{2}{h} \int T(E) (E-\mu)^n (-d/dE f(E,\mu,\Theta)) dE,
\label{eq:L_n}
\end{equation}
where $f(E,\mu,\Theta)$ is the Fermi occupation function at a given energy $E$, electrochemical potential $\mu$, and temperature $\Theta$. In linear response the conductance $G$ and the Seebeck coefficient $S$ are given by \cite{ouyang09}
\begin{equation}
G=e^2 L_0 \ \ \ \ \ \ S=-\frac{1}{e \Theta} \frac{L_1}{L_0}.
\label{eq:G_S}
\end{equation}
These quantities are calculated for parallel and anti-parallel magnetic orientation of the $Fe_xCo_{1-x}$ layers to eventually calculate the TMS ratio according to Eq.~(\ref{eq:TMS}). Note, that the conductance is basically the area under the transmission function times the derivative of the occupation function, whereas the Seebeck coefficient is proportional to the expected value (or first moment) of the very same product. Therefore, the Seebeck coefficient is determined by the asymmetry of the transmission function with respect to the Fermi level. Normally, the geometric mean of a function is more affected by small modulations of the function in comparison to the area under that function. Consequently, it can be expected that the Seebeck coefficient can depend crucially on changes of the transmission function e.g. due to alloying of the magnetic material. 
All calculations are done in the atomic sphere approximation and the cut-off for the angular momentum is 3. The k-point grids consist of 576 and 40,000 points in the whole Brillouin zone for the self-consistent and transport calculation, respectively.

\begin{figure}
\includegraphics[width=0.95 \linewidth]{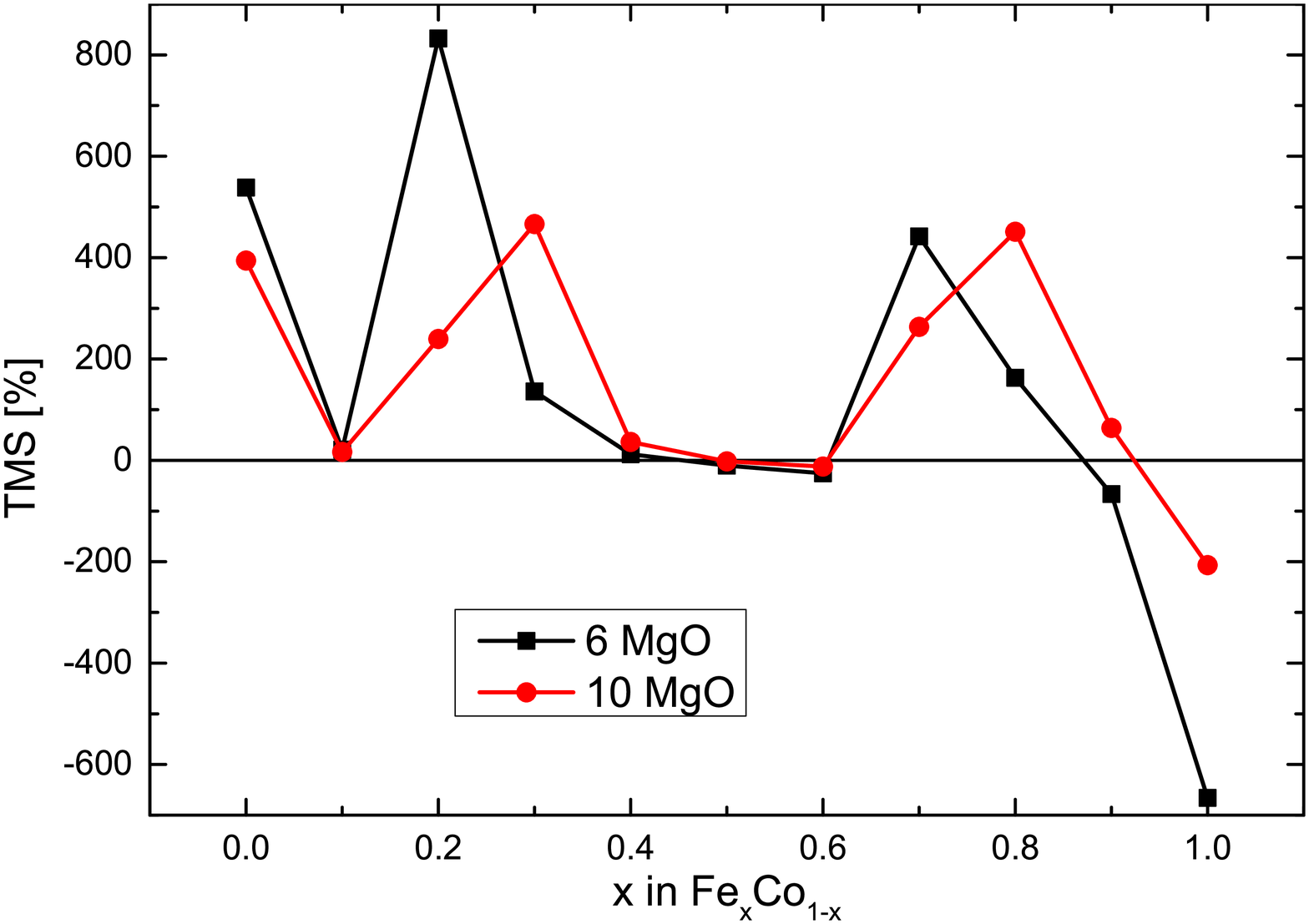}
\includegraphics[width=0.95 \linewidth]{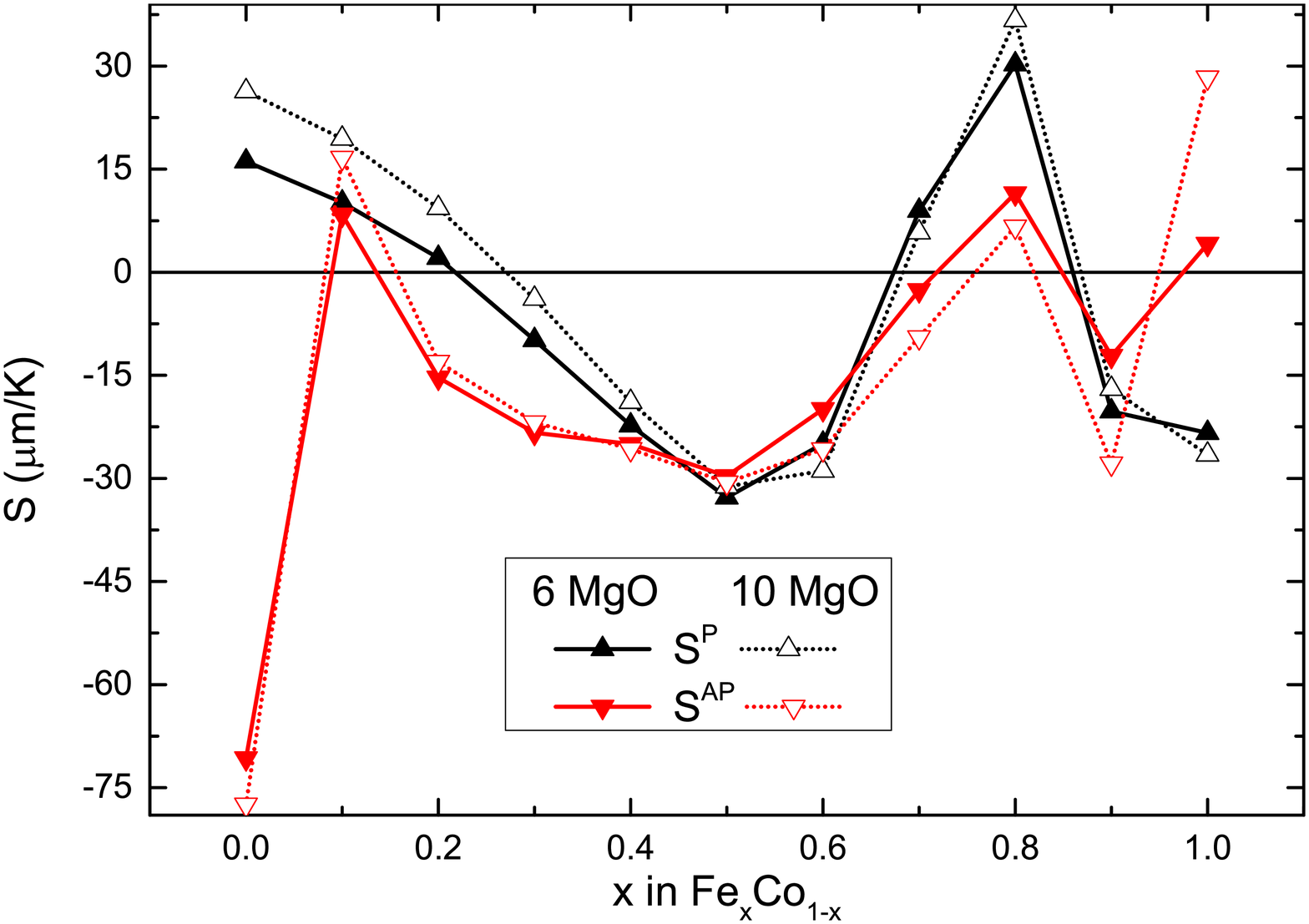}
\caption{
(Color online) Top: Tunneling magneto-Seebeck (TMS) ratio as a function of $Fe_xCo_{1-x}$ composition at room temperature for two different barrier thicknesses. Bottom: Seebeck coefficient for parallel $S^P$ and anti-parallel $S^{AP}$ magnetic configuration of the ferromagnetic layers as a function of the $Fe_xCo_{1-x}$ composition at room temperature for two different barrier thicknesses.
}
\label{comp}
\end{figure}

One of our main results is shown in Fig.~\ref{comp}. There, the TMS ratio at room temperature for two different barrier thicknesses is given as a function of the $Fe_xCo_{1-x}$ composition. It clearly shows that not only the magnitude but also the sign of the TMS ratio is very sensitive to the actual composition. This means that even small changes in the composition can drastically change the magnitude and even the sign of the TMS ratio. This could explain the small experimentally observed values~\cite{walter11,liebing11} in comparison to the large values predicted for the pure materials~\cite{czerner11}, because in experiments usually compositions close to fifty-fifty $Fe$ and $Co$ are used. In addition, differences in the composition of the magnetic material can be a reason for the different signs observed in the experiments in Ref.~\onlinecite{walter11} and Ref.~\onlinecite{liebing11}.

Increasing the barrier thickness from six to ten monolayers can change the values significantly depending on the composition. In particular, the TMS ratio can increase as well as decrease with increasing barrier thickness and even the sign can change. To analyze the TMS further the bottom viewgraph of Fig.~\ref{comp} shows the Seebeck coefficients at room temperature for parallel and anti-parallel magnetic configuration that leads to the discussed TMS ratios according to Eq.~(\ref{eq:TMS}). Likewise, here no general trend is visible and the Seebeck coefficient can be decreased or increased by increasing the barrier thickness.

\begin{figure*}
\includegraphics[width=0.45 \linewidth]{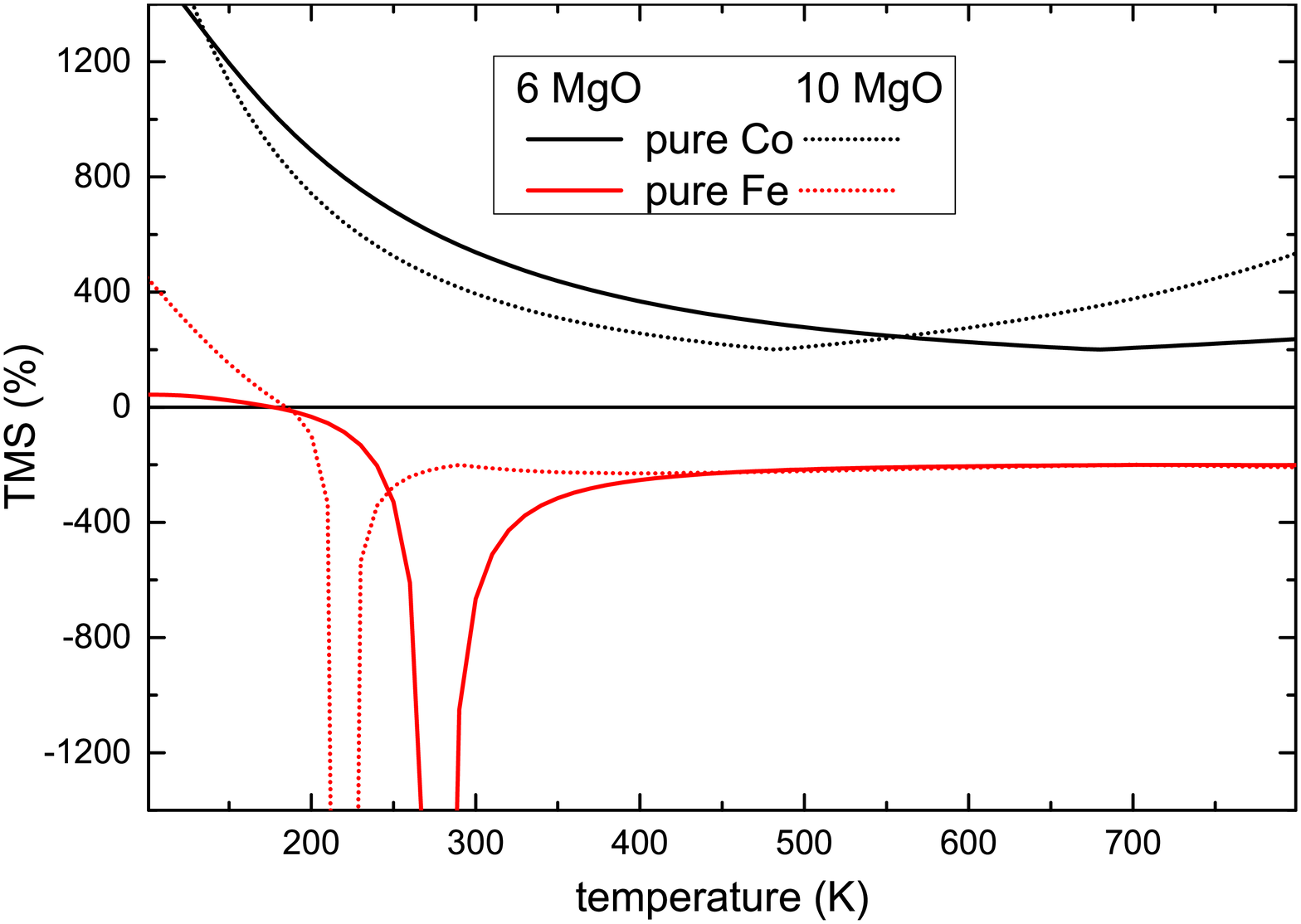}
\includegraphics[width=0.45 \linewidth]{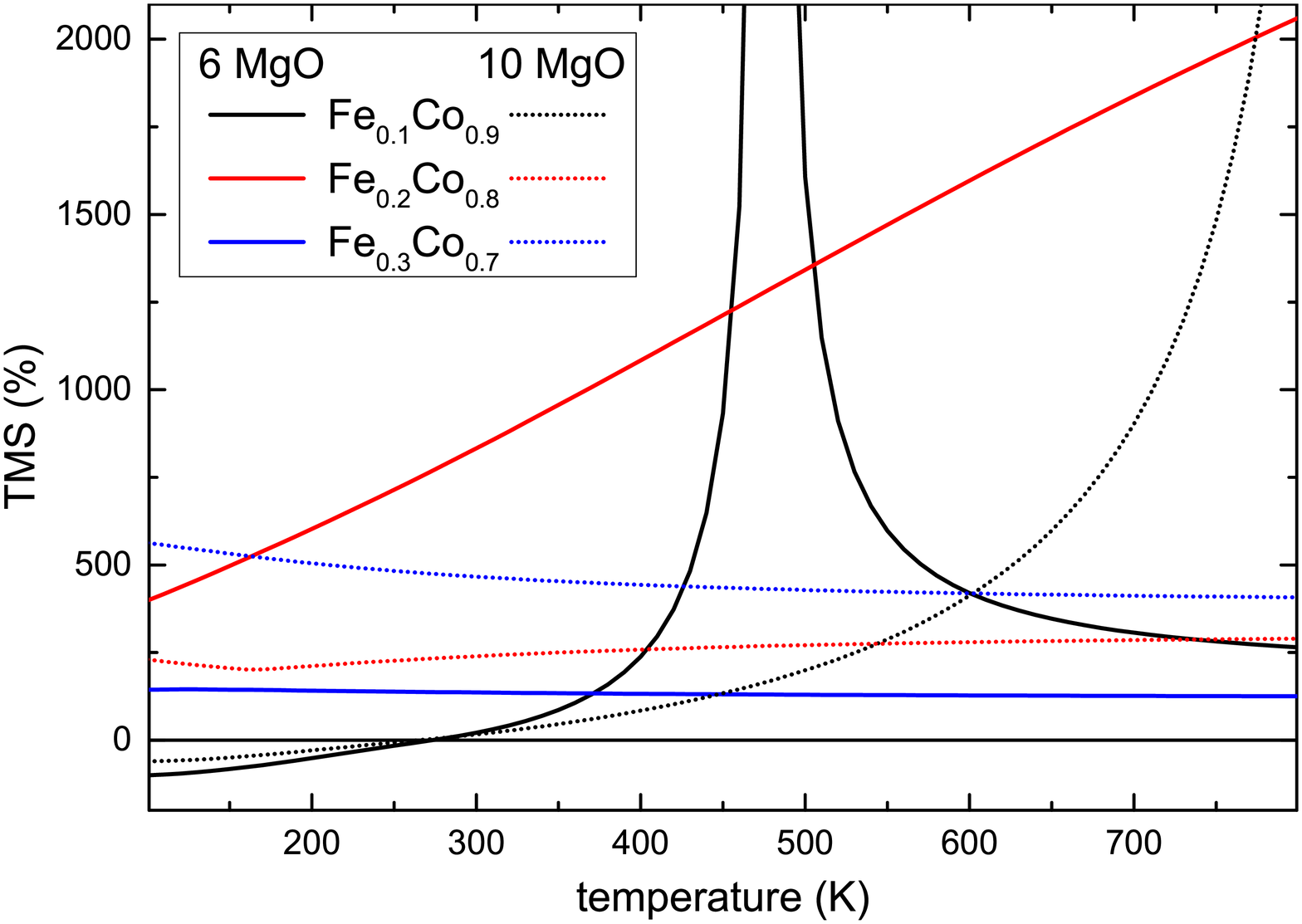}
\includegraphics[width=0.45 \linewidth]{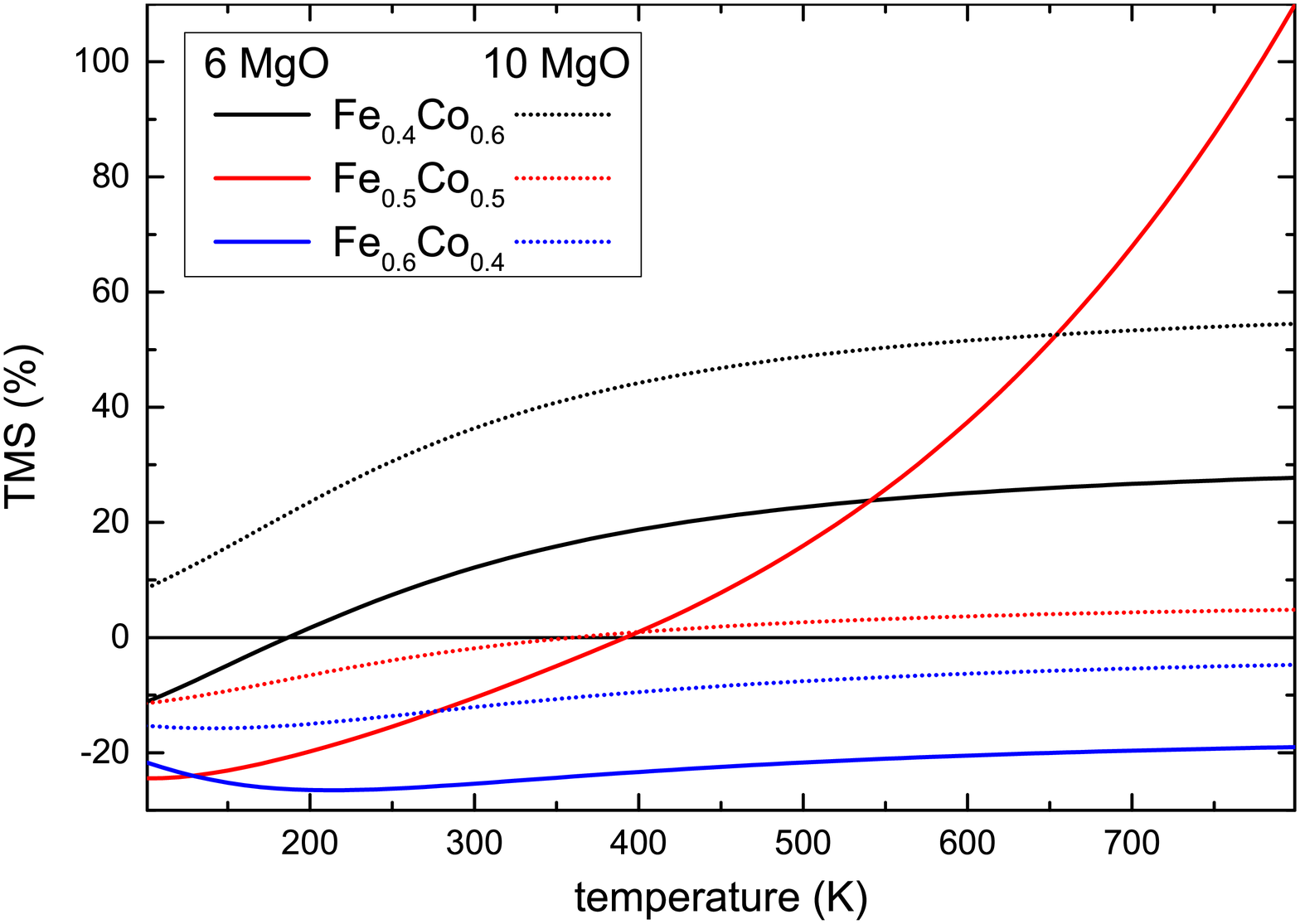}
\includegraphics[width=0.45 \linewidth]{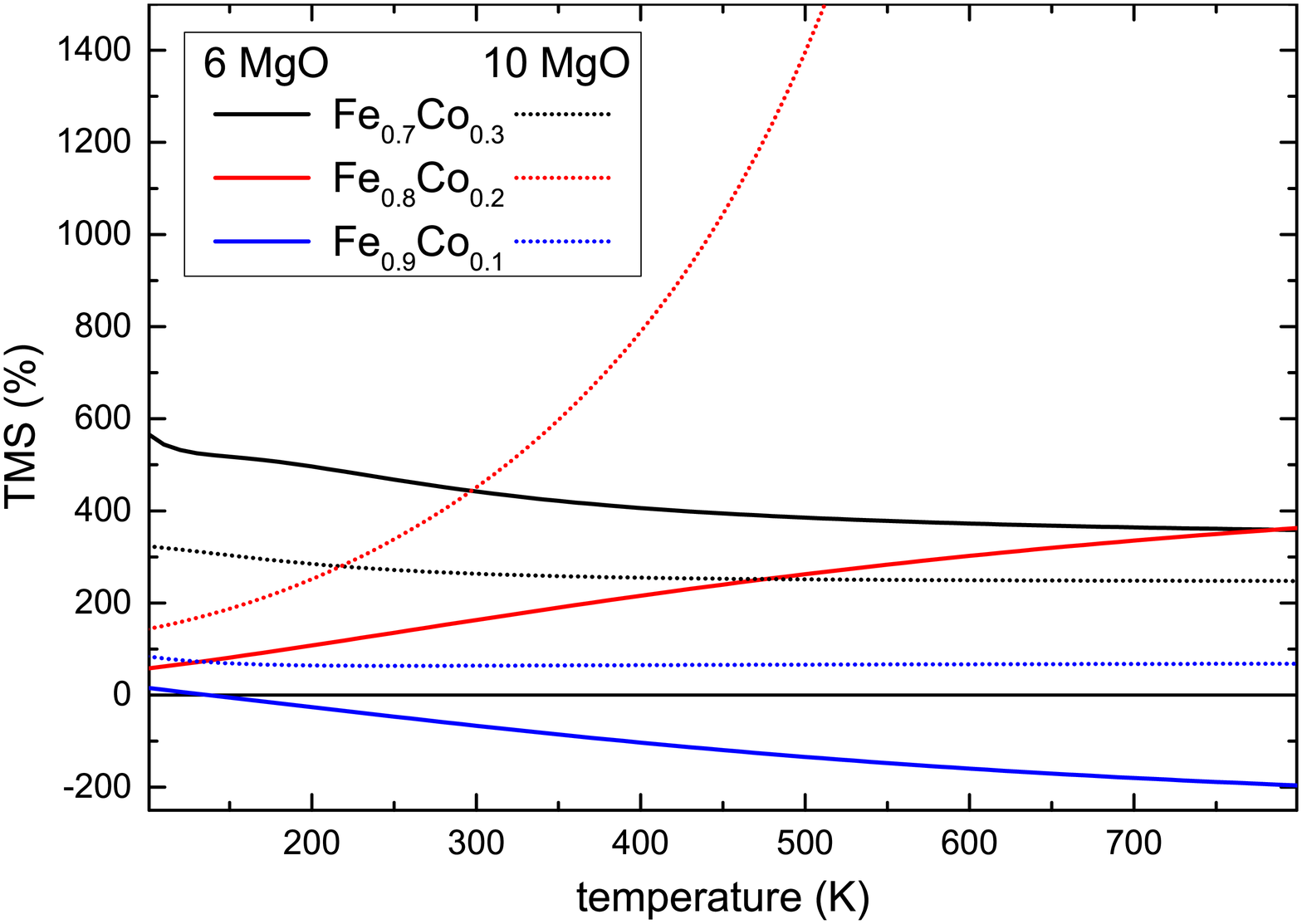}
\caption{
(Color online) Tunneling magneto-Seebeck (TMS) ratio as a function of temperature for pure Fe and Co (as already published in Ref.~\onlinecite{czerner11}) and for different alloy compositions of $Fe_xCo_{1-x}$. For the barrier we use six and ten monolayers of $MgO$.
}
\label{TMS}
\end{figure*}

Up to now we discussed the TMS only at room temperature. In Fig.~\ref{TMS} we show for the different considered compositions the temperature dependence of the TMS ratio for six and ten monolayers of $MgO$. Note, that the temperature enters the calculation only in the occupation function in Eq.~(\ref{eq:L_n}). The temperature dependencies for the pure materials were already discussed in Ref.~\onlinecite{czerner11}. Fig.~\ref{TMS} shows that the temperature dependence is rather involved and that there is also no clear trend visible when the composition is changed. We illustrate this with two examples. First, in the lower left panel all shown compositions have a moderate increase except for the $Fe_{0.5}Co_{0.5}$ alloy, which has a strong increasing TMS ratio with temperature. Second, in the lower right panel for $Fe_{0.7}Co_{0.3}$ and $Fe_{0.9}Co_{0.1}$ the TMS ratio is decreasing with increasing temperature but for the composition in between $Fe_{0.8}Co_{0.2}$ the TMS ratio is increasing. Besides the dependence on the composition also the change in the barrier thickness can lead to quite different temperature dependencies, for example visible for $Fe_{0.2}Co_{0.8}$, $Fe_{0.5}Co_{0.5}$, and $Fe_{0.8}Co_{0.2}$.

Particularly striking features are divergences of the TMS ratio which occur when one of the Seebeck coefficients goes through zero. Besides for pure $Fe$ only for the $Fe_{0.1}Co_{0.9}$ alloy a divergence is observed. Another interesting feature is a sign change of the TMS ratio with temperature, which is also observed experimentally~\cite{walter11}. To get an overview of when the sign changes occur we compile in Tab.~\ref{tab:sign} the temperatures at which the sign changes occur and in which direction it proceeds. For the compositions where no sign change occurs we give the corresponding sign of the TMS ratio over the whole temperature range. We see that there is a sign change of the TMS ratio only for certain compositions and that it also depends on the barrier thickness. Moreover, a comparison of $Fe_{0.6}Co_{0.4}$ and $Fe_{0.7}Co_{0.3}$ shows again that the sign of the TMS ratio crucially depends on the composition of the ferromagnetic material.

\begin{table*}
	\centering
		\begin{tabular}{cc||p{1.1cm}|p{1.1cm}|p{1.1cm}|p{1.1cm}|p{1.1cm}|p{1.1cm}|p{1.1cm}|p{1.1cm}|p{1.1cm}|p{1.1cm}|p{1.1cm}}
		& $x$ in $Fe_xCo_{1-x}$ & 0.0 & 0.1 & 0.2 & 0.3 & 0.4 & 0.5 & 0.6 & 0.7 & 0.8 & 0.9 & 1.0 \\
		\hline
		\hline
		6 MgO & $T_{sc} (K)$ & - & 285 & - & - & 185 & 390 & - & - & - & 140 & 175 \\
		& & $p$ & $n \rightarrow p$ & $p$ & $p$ & $n \rightarrow p$ &  $n \rightarrow p$ & $n$ & $p$ & $p$ & $p \rightarrow n$ &  $p \rightarrow n$ \\
    \hline
		10 MgO & $T_{sc} (K)$ & - & 265 & - & - & - & 360 & - & - & - & - & 185 \\
		& & $p$ & $n \rightarrow p$ & $p$ & $p$ & $p$ &  $n \rightarrow p$ & $n$ & $p$ & $p$ & $p$ &  $p \rightarrow n$
		\end{tabular}
	\caption{Temperatures of the sign changes $T_{sc}$ of the tunneling magneto-Seebeck (TMS) ratio, if present, extracted from the different temperature dependencies given in Fig.~\ref{TMS}. In addition, the direction of the sign change is given in the table: $n \rightarrow p$ ($p \rightarrow n$) means a change from negative to positive (positive to negative) with increasing temperature. If no sign change is present $n$ or $p$ gives the sign of the TMS ratio over the whole temperature range.    }
	\label{tab:sign}
\end{table*}

\begin{figure*}
\includegraphics[width=0.99 \linewidth]{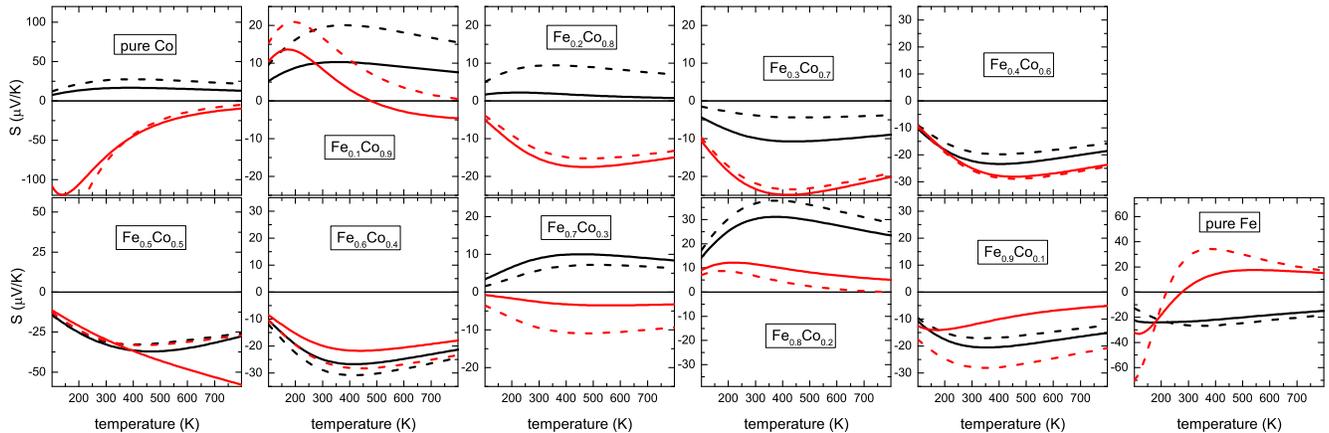}
\caption{
(Color online) Seebeck coefficients for parallel (black) and anti-parallel (red) magnetic configuration for different $Fe_xCo_{1-x}$ compositions. For the barrier we use six (solid) and ten (dashed) monolayers of $MgO$.
}
\label{seebeck}
\end{figure*}

For completeness, we give in Fig.~\ref{seebeck} the corresponding temperature dependencies of the Seebeck coefficients for parallel and anti-parallel magnetic configuration, which lead to the temperature dependencies of the TMS shown in Fig.~\ref{TMS}. It shows that a vanishing Seebeck coefficient in the anti-parallel configuration is responsible for all the observed divergencies, which is just by accident and has no physical reason.

\begin{figure}
\includegraphics[width=0.99 \linewidth]{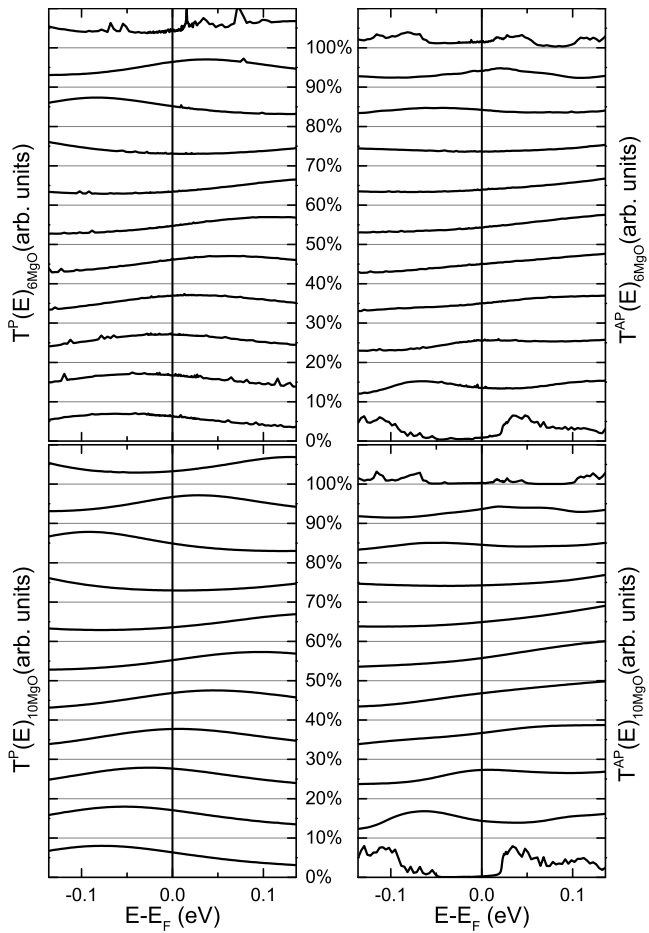}
\caption{
Transmission function $T(E)$ for parallel (left) and anti-parallel (right) magnetic configuration for different $Fe$ content in $Fe_xCo_{1-x}$. For the barrier we use six (top) and ten (bottom) monolayers of $MgO$. Note, that within each of the four panels the same scale of $T$ is used. The scale of $T^{AP}$ is $0.05$ ($0.03$) times the corresponding scale of $T^P$ for six (ten) monolayers of $MgO$, and the scale of $T^P$ for ten monolayers of $MgO$ is $0.0015$ times the scale of $T^P$ for six monolayers of $MgO$. 
}
\label{trans}
\end{figure}

Up to now, all the presented results seem unsystematic with respect to the alloy composition. The underlying physical quantity is the transmission function $T(E)$, which determines the transport properties according to Eqs.~(\ref{eq:L_n}) and (\ref{eq:G_S}). Consequently, in Fig.~\ref{trans} we look at $T(E)$ for the parallel and anti-parallel configuration for the two considered barrier thicknesses. For the transmission in the parallel configuration (left panels in Fig.~\ref{trans}) a systematic and continuous change with the ferromagnetic layer composition is visible. For example, for six monolayers of MgO there is a peak in $T^P(E)$ below the Fermi level, which moves continuously to higher energies with increasing $Fe$ concentration. For an $Fe$ concentration of about 70\% a second peak occurs at energies below the Fermi level. One has to keep in mind that the Seebeck coefficient is proportional to the first moment of the transmission function times the derivative of the occupation function (see Eqs.~(\ref{eq:L_n}) and (\ref{eq:G_S})). Consequently, only the asymmetry of $T(E)$ is important and this is changed by the discussed movement of peaks. With this the given temperature dependencies of $S^P$ in Fig.~\ref{seebeck} can be understood. Moreover, due to the continuous change of $T^P(E)$ one can interpolate to other compositions without doing a full \textit{ab initio} calculation.

The reason for this continuous change in the parallel configuration can be explained with the band structure. First, it is important to realize that the transport is dominated by the majority spin in $Fe$ and $Co$ where a $\Delta_1$ band is present and dominates the transport~\cite{butler01,heiliger06,heiliger08b}. By alloying this band is only weakly affected and the composition primarily shifts the position of the Fermi level. And this shift of the Fermi level corresponds to the shift of the transmission function with the composition.

However, this is no longer the case for the anti-parallel configuration. Here, the simple $\Delta_1$ band does not contribute to the transport and the transport is dominated by pockets within the Brillouin zone~\cite{butler01}. This is clearly visible in the rich structure of $T^{AP}$ for the pure materials. In the alloy case the band structure is broadened by disorder-scattering leading to a "washed out" transmission function. Going through the different compositions no clear continuous change is visible, in particular, close to the pure materials. Therefore, the rather involved temperature dependencies of the TMS ratio can be traced back to the complicated change with composition of the transmission function in the anti-parallel magnetic configuration.

In conclusion, we show that the TMS ratio is crucially dependent on the alloy composition of the magnetic material. The behavior seems unsystematic but can be traced back to a simple concentration dependence of the parallel transmission and a complicated change of the transmission function in the anti-parallel magnetic configuration. This leads to a strong dependence not only of the magnitude but also of the sign of the TMS ratio even for small changes in compositions. In general, the TMS ratios are smaller for compositions close to fifty-fifty in comparison to compositions closer to the pure materials. Moreover, small changes in compositions can cause the different signs of the TMS, which were observed in experiments~\cite{walter11,liebing11}. To achieve a full understanding of the TMS in the $MgO$ based tunnel junctions further experimental and theoretical studies are necessary.

We thank M. M\"unzenberg and M. Walter for useful discussions and acknowledge support from DFG SPP 1386 and DFG grant HE 5922/1-1.

%
%
%


\end{document}